\documentclass{aa}  

\usepackage{graphicx}
\usepackage{txfonts}
\usepackage{amsmath}
\usepackage[colorlinks=true, citecolor = blue]{hyperref}
\usepackage{subfigure}
\begin{document} 
   \title{The need for spatially resolved observations of PAHs in protoplanetary discs}
   \titlerunning{The need for spatially resolved PAH emission} 
   \subtitle{}
   \author{K. Lange
          \inst{1},
          C. Dominik
          \inst{1}
          \and
          A. G. G. M. Tielens
          \inst{2,}\inst{3}}

   \institute{Anton Pannekoek Institute for Astronomy, University of Amsterdam,
              Science-Park 904, 1098 XH Amsterdam, Netherlands;
              \email{k.lange@uva.nl}
         \and
             {Leiden Observatory, Leiden University, P.O. Box 9513, 2300 RA Leiden, Netherlands}
        \and {Astronomy Department, University of Maryland, MD 20742, USA}
             }

   \date{Received xxx; accepted xxx}

  \abstract
    {The signatures of polycyclic aromatic hydrocarbons (PAHs) have been observed in protoplanetary discs, and their emission features obtained from spectral energy distributions (SED) have been used in the literature to characterise their size and determine their abundance.}
    {Two simple disc models (uniform PAH distribution against a PAH gap in the inner disc) are compared to investigate the difference of their SED and obtainable information.}
    {We used the radiative transfer code RADMC-3D to model the SED of two protoplanetary discs orbiting a typical Herbig star, one of which features a depletion of PAHs in the inner disc. We further created artificial images of the discs at face-on view to extract radial profiles of the PAH emission in the infrared.}
    {We find that the extracted PAH features from an SED provide limited information about the PAHs in protoplanetary disc environments, except for the ionisation state. The distribution of PAHs in a protoplanetary disc influences the total observed PAH luminosity in a non-linear fashion and alters the relative strength between the 3.3\,$\mu$m and 11.3\,$\mu$m features. Furthermore, we produced radial profiles at the 3\,$\mu$m, 6\,$\mu$m and, 11\,$\mu$m PAH emission features and find that they follow a double power-law profile where the slope reflects the radiative environment (single photon regime vs. multi-photon regime) in which the PAHs lie.}   
    {Using spatially resolved techniques such as IFU or imaging in the era of the James Webb Space Telescope, we find that multi-wavelength radial emission profiles will not only provide information on the spatial distribution of the PAHs, but may also provide information on their size and underlying UV environment, which is crucial for photo-evaporative disc wind models.}


   \maketitle


\section{Introduction}
Polycyclic aromatic hydrocarbons (PAHs) are a group of organic molecules that contain two or more aromatic carbon rings.
These molecules have been observationally confirmed in the interstellar medium (ISM), galaxies, post asymptotic giant branch (post-AGB) stars and protoplanetary discs \citep[e.g.][]{Allamandola1985, Allamandola1989, Meeus2001, Kennicutt2003, Geers2006, Acke2010} through their distinct infrared emission features between 3-20\,$\mu$m obtained with ground-based and space-based observatories such as VISIR on the Very Large Telescope (VLT), the Spitzer Space Telescope, and the Infrared Observatory (ISO).
Future missions such as Ariel and Twinkle will be also capable of detecting PAH emission features in the infrared in protoplanetary discs and in principle also in exoplanetary atmospheres \citep{Ercolano2022}.\\
\\Surveys in the past have searched systematically for these molecules in planet-forming discs \citep[i.a.][]{Habart2004}.
\citet{Geers2006} searched amongst T\,Tauri stars for signs of PAHs, finding that only 8\,\% of the T\,Tauri discs show emission features, excluding tentative detections.
Recently, \citet{Valegard2021} found a 27\,\% detection rate of PAHs in intermediate-mass T\,Tauri star discs (IMMT) through visual inspection of the spectrum, excluding a few tentative detections as well.
\citet{Acke2010} searched for PAH features in Herbig Ae/Be discs and found a detection rate of PAH features of about 60\,\%.
Despite a clear relation between effective stellar temperature and the PAH detection rate that can most likely be traced back to the weaker UV fields in fainter and colder systems, it is not well understood what causes the difference in PAH emission between two sources of similar stellar type and why some sources are lacking emission at all.\\
\\Most of the available data have been obtained by the Spitzer Space Telescope using the infrared spectrograph (IRS) between 5\,$\mu$m and 40\,$\mu$m. 
Unfortunately, data for the very distinct 3.3\,$\mu$m PAH feature are only available for a few sources obtained by the ISO, SWS, ISOPHOT or VLT ISAAC \citep[see e.g.][for an overview of available data]{Seok2017}.
One year ago, the James Webb Space Telescope (JWST) launched, carrying the near-infrared spectrograph (NIRSpec), the near infrared camera (NIRcam) and the mid-infrared instrument (MIRI). They allow obtaining high-quality data of the important PAH features at 3.3\,$\mu$m, 6.2$\,\mu$m, 7.7$\,\mu$m, 8.6\,$\mu$m, and 11.3\,$\mu$m  with a single facility.
These instruments are well-suited for characterising the PAH content in protoplanetary discs, especially with the NIRSpec and MIRI integral field units, which for the first time allow us to obtain spatially resolved spectra of PAH emission especially aimed at T\,Tauri discs.
The JWST is expected to provide an unbiased and complete overview of the spatial and spectral characteristics of PAHs in protoplanetary discs.\\
\\To guide JWST observations and interpret the amount of newly available data in the coming years, it is necessary to make models so that the upcoming results can be properly interpreted.
For this purpose, we created two disc models that contain PAHs and performed radiative transfer calculations to investigate two different aspects of the upcoming observations.
First, we wish to analyse the spectra of protoplanetary discs seen under different inclinations and characterise the different PAH feature strengths, their origin, and the possible interpretations of upcoming new PAH detections.
Second, both NIRSpec and MIRI on board the JWST can provide integral field unit spectroscopy at unpreceeded resolution and sensitivity, allowing us to extract radial profiles of the important PAH features at 3.3\,$\mu$m, 6.2$\,\mu$m, and 11.3\,$\mu$m. 
Therefore, we wish to provide calculated profiles and predictions for the variation in the relative feature strengths between the three PAH features as a function of the distance from the host star.\\
\\This work is structured in the following way: first, we explain our disc model and our setup for the radiative transfer code RADMC-3D. Then, we present the results we obtained for the SEDs under different inclinations, followed by a discussion of the radial profiles that we expect from a (dust)-gapless disc. Finally, we discuss and summarise our results.


\section{Methods}
In order to simulate the PAHs in a protoplanetary disc, we require a disc model, a radiative transfer model that determines the radiation field and temperature structure of the disc, and a model for the PAH emission itself.
For the radiative transfer model, we used the commonly used code RADMC-3D\footnote{\url{https://www.ita.uni-heidelberg.de/~dullemond/software/radmc-3d/index.php}} \citep{Dullemond2012}, which  can simulate the continuum radiative transfer of dust populations in protoplanetary disc environments, for example.
It uses a Monte Carlo method to calculate the radiative equilibrium temperature in each grid cell.
Photon packages that originate from heating sources such as a central star are distributed throughout the grid, where they can be absorbed and re-emitted at a different wavelength or scattered into a different direction. 
Then, if the temperature structure of the disc is known, RADMC-3D uses ray-tracing in order to obtain images and spectra from a certain inclination and distance of the observer to the object.
However, RADMC-3D does not self-consistently compute the gas temperature in a disc as it does not feature detailed photochemistry.
Hence, as PAHs and the stochastically heating and cooling of PAHs is not included in RADMC-3D itself, we are required to perform these calculations outside RADMC-3D and supply them later for the radiative transfer.
Therefore, we separate this section into two parts: the disc model and setup that we used for RADMC-3D to obtain the radiation field in the UV, dust grain temperatures, and the obtained continuum emission.
In the second part, we calculate the stochastically heating of the PAHs for different UV fields and the corresponding temperature distributions in order to calculate the PAH emission spectrum at the corresponding UV field in the disc.

\subsection{Disc model}
We focus on a smooth protoplanetary disc (no dust substructures as seen e.g. with ALMA in the DSHARP survey, \cite{Andrews2018}) in order to link our studies to earlier studies on PAHs and because our aim is to define a general framework rather than modelling a specific disc and its properties \citep[for the effect of dust gaps on PAH emission features see][]{Maaskant2014}.
We chose to model a disc around a Herbig star, where we take HD\,169142 as a reference model for the typical Herbig star.
However, these models can be readily extended to T\,Tauri stars, even though their UV radiation field is much weaker than in a typical Herbig star disc and additional photo-chemical processes of importance such as consideration of highly energetic UV and X-ray photons \citep{Siebenmorgen2010} might be necessary.\\
\\The gas in our disc follows the vertical equilibrium, where the density profile is given by
\begin{equation}
    \rho(z,r) = \rho_\text{mid}(r)\,\text{exp}\left(\frac{-z^2}{2H^2}\right),
\end{equation}
where $\rho_\text{mid}(r)$ marks the midplane density that is linked to the gas pressure scale height $H$ and the surface density $\Sigma(r)$ through $\rho_\text{mid}(r) = \Sigma(r)/\sqrt{2\pi}H(r)$.
The gas-pressure scale height is defined as $H = c_\text{s}/\Omega_\text{K}$ , with the speed of sound $c_\text{s}$ and the Keplerian frequency $\Omega_\text{K}$.
The speed of sound is given by $c_\text{s}=\sqrt{k_\text{B}T/\mu m_\text{p}}$ , with the Boltzmann constant $k_\text{B}$ and the mean molecular weight of the gas ($\mu=1.37$).
Our disc follows a gas surface density profile scaled to a minimum mass solar nebula (MMSN),
\begin{equation}
    \Sigma(r) = 730 \frac{\text{g}}{\text{cm$^2$}} \left(\frac{r}{1\,\text{au}}\right)^{-1.5} \left(\frac{M_*}{M_\odot}\right),
\end{equation}
as described by \citet{Weidenschilling1977}, so that the total mass of the disc is $M_\text{disc} = 0.01$\,M$_*$.
Our temperature profile is defined through the commonly used \citet{Hayashi1981} profile
\begin{equation}
    T(r) = 280\,\text{K} \left(\frac{r}{\text{au}}\right)^{-0.5}\left(\frac{L_*}{L_\odot}\right)^{0.25} \text{.}
    \label{eq:T_dust}
\end{equation}
For the disc and stellar parameters, we chose the inner disc radius to be $r_\text{in}=0.5$\,au, the outer disc radius as $r_\text{out}=100$\,au, the stellar mass $M_*=2.28$\,$M_\odot$, the stellar luminosity $L_*=15.33$\,$L_\odot$, and the stellar temperature $T_*=8200$\,K \citep{Honda2012}.
For the dust grains, we chose a minimum grain size of $a_\text{min} = 0.05$\,$\mu$m and a maximum grain size of $a_\text{max} = 1$\,mm following a Mathis-Rumpl-Nordsieck
(MRN) size distribution \citep{Mathis1977},
\begin{equation}
    n(a) \propto a^{-\gamma},
\end{equation}
where $\gamma =3.5$.
Our disc is assumed to be a settled disc in which grains are in an equilibrium state between vertical settling and vertical turbulence.
We followed the model of \cite{Fromang2007} to describe the vertical distribution of the grains, 
\begin{equation}
    \frac{\partial\rho_\text{d}}{\partial t} - \frac{\partial}{\partial z}\left(z \Omega^2 \tau_\text{s}\rho_\text{d}\right) = \frac{\partial}{\partial z} \left[D\rho \frac{\partial}{\partial z}\left(\frac{\rho_\text{d}}{\rho}\right)\right],
\end{equation}
where we obtained the steady-state solution by integrating the differential equation to solve for the dust density $\rho_\text{d}$,
\begin{equation}
    \frac{\partial}{\partial z}\left(\text{ln}\frac{\rho_\text{d}}{\rho}\right) = - \frac{\Omega^2 \tau_\text{s}}{D} z \text{.}
    \label{eq:dust}
\end{equation}
We used the standard relations provided by \citet{Fromang2007} for the diffusion coefficient $D$ and stopping time $\tau_\text{s}$, which were also used and described in \citet{Lange2023}.

\subsection{Stochastic heating of PAHs}
The PAHs are large aromatic molecules that are efficient in absorbing UV photons and emitting the gained energy as spectral features in the near-infrared (NIR) and mid-infrared (MIR).
As PAHs are individual molecules, their heat capacity is small, and the absorption of a single UV photon can raise their electronic excitation temperature by several hundred Kelvin \citep{Tielens2008}. 
As a consequence, the excitation temperature of PAHs is not well described by the radiative equilibrium temperature like for dust grains, but requires a stochastic treatment of the heating and cooling process.
Instead of describing the temperature only through the equilibrium value, the temperature probability distribution $G(T)$ needs to be determined to properly treat and model the emission of spectral features \citep{Bakes2001}. 
For a low photon absorption rate ($\leq 1$\,s$^{-1}$), the temperature probability distribution of the PAHs is dominated by the absorption of a single photon heating the PAH to a peak temperature, followed by a phase of cooling down to the background temperature.
For a high UV photon flux without consideration of photochemical processes that require energy such as photodissociation, the PAH will not have enough time to cool down to the background equilibrium temperature and will fluctuate around a mean temperature value in a broad Gaussian-like distribution \citep[for details and limits of this Gaussian distribution, see][]{Tielens2021, Lange2021}.
We evaluated the temperature distribution function for PAHs irradiated by a given UV field using a Monte Carlo method that accounts for the stochastic radiative cooling and heating processes.\\
\\We modelled the heat capacity of the PAHs by the microcanonical temperature-energy relation\footnote{As in \citet{Lange2021}, we smoothed the transition between the low- and high-temperature regime to preserve differentiability, $T_\text{m}=(T_\text{m,1}^8+T_\text{m,2}^8)^{0.125}$.} \citep{Tielens2021},
\begin{equation}
    T_\text{m} = \begin{cases}
    3750 \left(\frac{E\text{(eV)}}{3N-6}\right)^{0.45}\text{\,K}\text{\hspace{1cm} if $T_m < 1000$\,K}\\
    11000 \left(\frac{E\text{(eV)}}{3N-6}\right)^{0.8}\text{\,K}\text{\hspace{1cm} if $1000\text{\,K} < T_m$},\\
    \end{cases}
    \label{eq:T(E)}
\end{equation}
where $N$ describes the number of atoms in a PAH molecule, $T_\text{m}$ is the microcanonical temperature, and $E$ is the excitation energy.
From this relation, it becomes clear that the smallest PAHs undergo the strongest temperature fluctuations and reach the highest peak temperatures.
The absorption rate of photons $r_{a}$ is determined by the stellar spectrum that we assumed for Herbig stars to be a black body at the given stellar effective temperature of the host star and the absorption cross section of the PAH.
The absorbed energy flux $\Phi_{\text{a},\text{E}}$ at energy $E$ is then
\begin{equation}
    \Phi_{\text{a},\text{E}} = F_\text{E} \sigma_\text{E}
,\end{equation}
where $F_\text{E}$ is the flux density at given $E$ and $\sigma_\text{E}$ is the absorption cross section of a PAH at a given photon energy.
We considered photons between 0.1-13.6\,eV and bin the photon distribution into 100 bins, so that the absorption rate of photons of a given bin $m$ with average energy $E_m$ is 
\begin{equation}
    r_{\text{a}, m} = \frac{1}{E_m} \int_{E_\text{m, min}}^{E_\text{m,max}} \Phi_{\text{a}, \text{E}} \text{d}E \text{.}
\end{equation}
As the ionisation of hydrogen significantly increases the optical depth above 13.6\,eV \citep{Ryter1996}, we did not account for higher-energy photons. 
By assuming an average photon energy in the infrared $E_\text{IR}$, we can determine the rate $r_\text{e}$ at which photons are emitted for cooling, 
\begin{equation}
    r_\text{e} = \frac{\text{d}E}{\text{d}t} E_\text{IR}^{-1} \text{,}
\end{equation}
if the cooling rate $\frac{\text{d}E}{\text{d}t}$ is known. Together with equation \eqref{eq:T(E)}, we applied the cooling model from \citet{Bakes2001} and \citet{Tielens2021} to determine the cooling rate,
\begin{equation}
    \frac{\text{d}T_m}{\text{d}t} = -1.1 \cdot 10^{-5}T_m^{2.53} \text{\,K/s}
    \label{eq:dTdt}
.\end{equation}
Then, in order to evaluate the statistical heating and cooling, we determined the total rate of events for a Monte Carlo evaluation,
\begin{equation}
    r = \sum_m r_{\text{a}, m} + r_\text{e} \text{\,.}
\end{equation}
Similar to \citet{Zsom2008}, we used an exponential distribution for the time between two events,
\begin{equation}
    f(\delta t) = r \,\text{exp}(-r\delta t)
\end{equation}
with $R = \{ r_\text{{a},m}$, $r_\text{e}\}$.
Then, the probability of an event $j$ with rate $r_j \in R$ can be described by
\begin{equation}
    P(j) = \frac{r_j}{r},
\end{equation}
and the event is randomly drawn with the given probabilities.
As a result, we received the fluctuating temperature of the PAHs at any time and calculated the temperature probability distribution $G(T)$.

\subsection{PAH emission model}
As the next step in our calculation, we need to calculate the PAH emission spectra from the temperature probability distribution $G(E)$.
We used the NASA AMES PAH IR spectral database, where theoretical PAH spectra computed with density functional theory (DFT) are available \citep{Boersma_2014, Bauschlicher_2018, Mattioda_2020}.
Because we focused our calculations on the general spectral features of PAHs rather than modelling a specific disc, we use dcoronene C$_{24}$H$_{12}$, circumoronene C$_{54}$H$_{18}$, and circumcircumcoronene C$_{96}$H$_{24}$ in our calculations in the neutral and cation state to cover the whole size range of astrophysically relevant PAH molecules.
The spectra of PAH clusters are available in the database, and we expect that PAHs will be present on dust grains as clusters with different sizes, ranging from a few cluster members up to a few hundred.
However, we chose to use the spectrum of individual PAH molecules for both frozen-out (hereafter also called adsorbed) and gas-phase PAHs.
When we compare the results for photodissociation and desorption of PAH clusters in protoplanetary discs \citep{Lange2021}, the timescale for photodissociation of clusters is much shorter than the timescale for desorption of the cluster from the dust grain.
Hence, we do not expect a large fraction of clusters to be present in the gas phase, but to be dominantly adsorbed.
Additionally, the selection of clusters available in the database does not cover all expected cluster sizes.\\
\\Using the temperature probability distribution $G(T)$ from our Monte Carlo model, we followed the approach by \cite{Bakes2001}, where the full emission profile of a temperature fluctuating PAH in the gas-phase is given by
\begin{equation}
    F_\text{PAH, g}(\nu) = n_\text{PAH, g} \sum_Z L(\nu, Z) f(Z) \int B(T, \nu) G(T) \,\text{d}T,
\end{equation}
where $n_\text{PAH, g}$ defines the number density of gas-phase PAHs, $L(v,Z)$ describes the emission profile of the PAH as a function of the charge state $Z$, $f$ gives the fraction of PAHs that have a charge state $Z$, and $B(T,\nu)$ is Planck's function.
The integral over the Planck function $B$ and temperature probability distribution $G(T)$ defines the time-averaged black-body emission function.
We calculated the PAH emission $L(v,Z)$ by considering all available transitions as Lorentzians with the standard full width at half maximum (FWHM) value of 15\,cm$^{-1}$ (as in the NASA AMES spectral database) and integrated cross sections that are given in the database\footnote{note the unit conversion from km/mol to cm$^2$Hz through the conversion factor $10^5 \cdot c / $N$_\text{A}$ using the speed of light and Avogadro's constant.}.\\
\\For frozen-out PAHs, we considered the temperature fluctuations to be negligible as the heat transfer from the PAH to grain occurs on a much faster second timescale than the IR emission of an IR photon \citet{Lange2023}.
Therefore, we assumed that the frozen-out PAHs have the same temperature as their hosting dust grains at radiative equilibrium.
Under the assumption that all adsorbed PAHs are neutral, the PAH emission spectrum is calculated by
\begin{equation}
    F(\nu)_\text{PAH,s} = n_\text{PAH,s} \, L(\nu, Z=0) \,B(T_\text{dust}, \nu),
\end{equation}
where $n_\text{PAH, s}$ is the number density of frozen-out PAHs on dust grains. Because we did not consider the destruction of PAHs, we ensured that $n_\text{PAH, s} + n_\text{PAH, g}$ always equaled the expected abundance of PAHs from the ISM.\\
\\Because we did not include the statistical heating and cooling of PAHs into RADMC-3D itself, we needed to parametrise our calculations to make them accessible to RADMC-3D.
For this purpose, RADMC-3D provides the function \emph{userdef\_srcalp()} to add fluxes to every grid cell without interfering with the initial calculations of the dust temperature and calculation of the local radiation field.
Unfortunately, this means that any contributions to the opacity for the whole disc model are not considered or must be taken into account separately in a non-consistent way.
However, because we only considered a PAH abundance of 1\,\% of the ISM value, the contribution of PAHs to the opacities is negligible.\\
\\We chose the local UV field characterised by $G_0$ to be our parametrisation for all calculations. 
For this purpose, we sampled the temperature fluctuations $G(T)$ of the gas-phase PAHs in a logarithmic grid between $1 \leq G_0 \leq 10^9$ with 40 grid points and interpolated then to the nearest neighbour.
Opacity effects by the dust were then considered by RADMC-3D during the image and SED calculation itself.\\
\\We used two disc models that we wished to compare in our studies.
Our \emph{smooth-disc} model has a gas-phase PAH abundance of 1\,\% of the ISM relative to the gas density \citep[$0.01 \,\cdot\, $C$_\text{PAH}$:H = $1.5\cdot10^{-5}$][]{Tielens2008}), where the 99\,\% of the ISM abundance are frozen out on grains.
Our \emph{PAH-gap} disc model has a gas-phase PAH abundance of only 0.1\,\% relative to the ISM in the inner disc ($r \leq 40$\,au) and a gas-phase PAH abundance 1\,\% relative to the ISM in the outer disc, where again the remaining PAHs are frozen out. 
We chose this model as a test model for the results of \citep{Lange2023}, where coagulation and adsorption can lead to a depletion of PAHs in the inner disc in a Herbig disc model.
In this way, we can test the effect on the observable PAH flux ratios and detectability with current facilities such as the JWST.

\subsection{Ionisation alance}
Furthermore, we took the charge in the PAHs into account in our emission model. 
To do this, we evaluated the PAH charge and followed the ideas of \citet{Maaskant2014, Bakes1994, Tielens2005}, and references therein. 
In their approach, the ionisation balance is set by the ionisation parameter
\begin{equation}
    \gamma = G_0 T^{1/2} n_\text{e},
\end{equation}
where $n_\text{e}$ is the free electron density.
Assuming that the photosphere behaves like a photodissociation region \citep{Woitke2009}, we assumed that the free electrons are supplied by ionised carbon C$^{+}$, and we assumed an ISM abundance of elemental carbon that is dominantly ionised in an UV environment, $n_\text{C} = n_\text{e} = 1.5 \cdot 10^{-4} n_0$.
Then, the fraction of neutral PAHs is determined by
\begin{equation}
    f = 1 / (1 + \gamma_0),
\end{equation}
where the ratio of ionisation to recombination is given through
\begin{equation}
    \gamma_0 = 3.5 \cdot 10^{-6}\,N_\text{C}\,\gamma,
\end{equation}
where the pre-factor orginates from the electron-sticking probability \citep{Tielens2005}, and $N_\text{C}$ is the number of carbon atoms of the PAH molecule.
This is a simplified approach, and a detailed model would require a complex disc chemistry framework to solve the charge balance in the different molecular species.

\begin{figure*}
    \centering
    \includegraphics[width=0.99\linewidth]{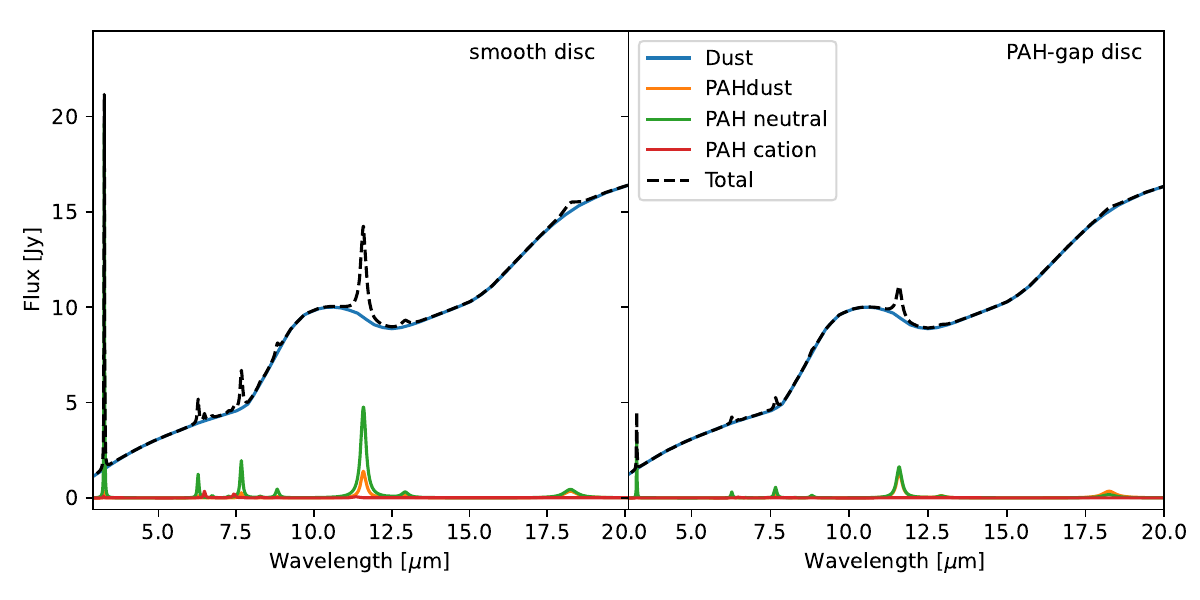}
    \caption{Simulated SED contributions of the \emph{smooth-disc} model (\emph{left}) and the \emph{PAH-gap} model (\emph{right}) seen in a face-on view. The PAH features are dominated by neutral emission, as expected \citep{Maaskant2014}. While the \emph{smooth-disc} model emission mainly comes from gas-phase PAHs, the \emph{PAH-gap} model has a significant contribution from the adsorbed PAHs that dominate the longer-wavelength features. The integrated PAH emission in both models does not scale with the total number of PAHs in the disc because five times more PAH flux is missing than in PAH molecules.}
    \label{fig:PAH-SED}
\end{figure*}

\subsection{Calculation sequence}
The setup and calculation of the models involves calculations that are performed outside RADMC-3D, but require information about the disc setup and the radiative transfer model itself.
To calculate the opacities of the dust grains, we used optool\footnote{\url{https://github.com/cdominik/optool}} \citet{Dominik2021}, a command line tool for calculating dust opacities for various mixtures of materials and grain size distributions. 
For our dust, we chose the standard DIANA opacities \citep{Woitke2016}.
The calculations were performed in the following order.
First we performed Monte Carlo calculations for the PAH temperature fluctuations in the range of the UV field $1 \leq G_0 \leq 10^9$ that we expect in the disc. Then, we used the PAH emission model to obtain the expected fluxes of each individual PAH molecule in the given UV radiation field.
Next, we set up the disc model itself in RADMC-3D and obtain the temperature structure of the disc with \emph{mctherm} and derived the local radiation field of each grid cell with \emph{mcmono} in the photon-energy range between 5\,eV and 13.6\,eV, and we calculated $G_0$ in each grid cell.
Based on $G_0$ and the number density of gas-phase PAHs in the grid cell, we calculated the PAH flux that is emitted in each grid cell.
Finally, we supplied RADMC-3D with the PAH fluxes in the function \emph{userdef\_srcalp()} and calculated images and spectra. For each of the images, we used $5\cdot10^{8}$ photons, and for the spectra, we used $10^{8}$ photons at each wavelength.


\begin{figure*}
    \centering
    \includegraphics[width=0.99\linewidth]{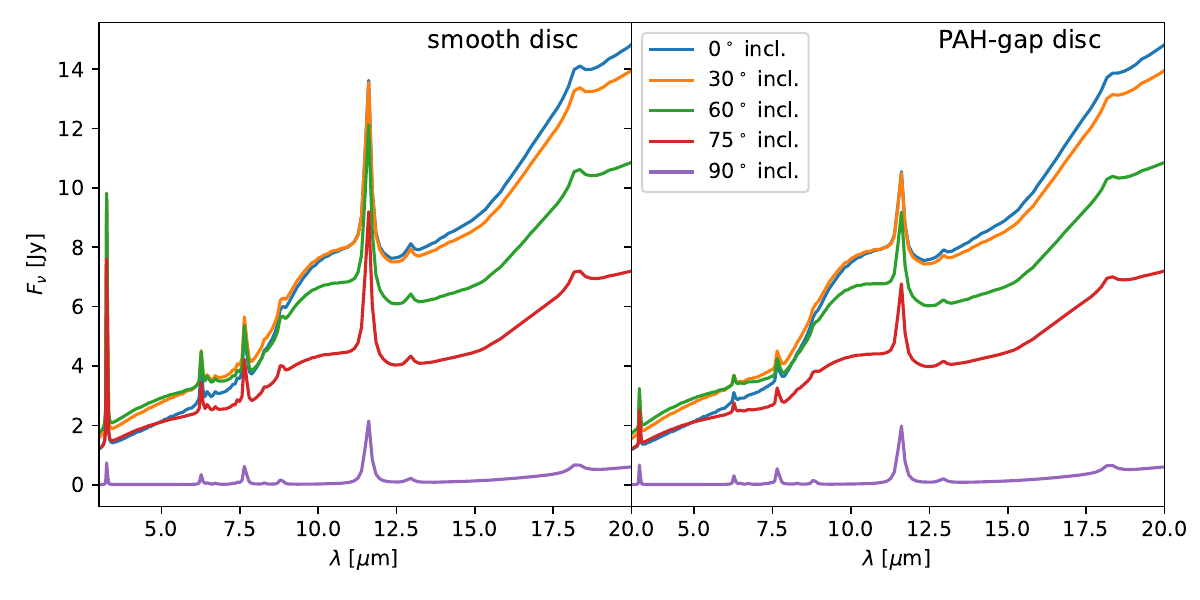}
    \caption{Simulated spectra of the two PAH disc models for selected inclinations. The left panel shows the SED of a smooth PAH disc, and the bottom panel shows the disc with a 40\,au PAH gap containing only 10\,\% PAHs in the gap compared to the smooth-disc model. Even though the strength of all PAH features decreases in the gap model, the short-wavelength features (3\,$\mu$m and 6\,$\mu$m) become much weaker than the 11\,$\mu$m feature.}
    \label{fig:PAH_SED_incl}
\end{figure*}

\begin{figure}
    \centering
    \includegraphics[width=0.99\linewidth]{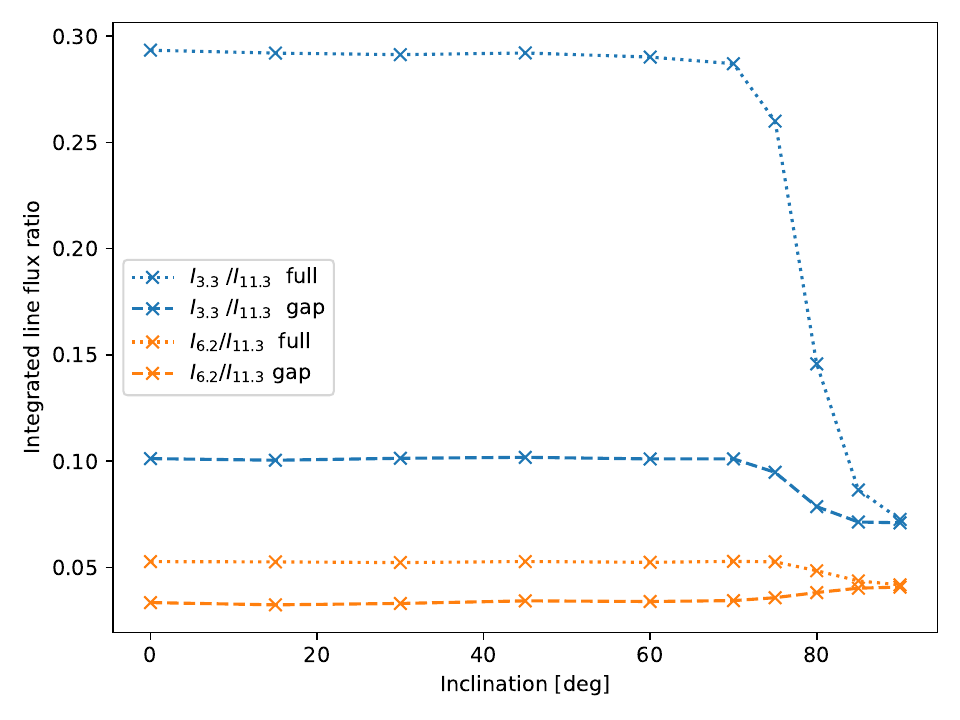}
    \caption{Intensity ratio of the 3.3\,$\mu$m and 6.2\,$\mu$m feature compared to the 11.3\,$\mu$m feature. The PAH gap in the inner disc clearly affects the observable feature ratio by a factor 1.6 to 3.}
    \label{fig:integrated_lineflux}
\end{figure}

\section{Results}
\subsection{PAH SEDs from protoplanetary discs}
We wish to analyse the expected PAH SEDs from protoplanetary discs in a systematic way. 
Therefore, we ran radiative transfer models with RADMC-3D to simulate observations of discs at different inclinations and locally different PAH abundances. 
We aim to understand the general effects and do not wish to reproduce the observational spectra obtained by \citet{Acke2010, Seok2017} for example, and we used the theoretical emission profiles of coronene (C$_{24}$H$_{12}$) from the NASA AMES PAH IR spectral database rather than the commonly used model by \citet{Draine2007}.
The model by \citet{Draine2007} was built to reproduce the shape of PAH features observed in protoplanetary discs by an average disc PAH spectrum.
It can be scaled to different sizes, including a transition to small carbonaceous grains.
However, the strength of the features is built to match general trends from observations and is only guided by calculated PAH spectra. 
The effect of the radiation environment in the disc (neutral to cation ratio, PAH sizes, and disc structures) on PAHs is therefore already implicitly built into the model by \citet{Draine2007}.
By using the NASA Ames database, we used only the intrinsic properties of PAHs and avoided these implicit assumptions about discs.\\
\\Furthermore, it is not possible to identify individual PAHs or to fit features of specific substituents (e.g. replace N or O with C) with it.
Instead, advanced models including many PAH species should be developed that not only reproduce the observations, but also allow conclusions about the underlying chemistry and radiation field.
We wish to motivate the usage of such a model by applying single PAH spectra from the database to a radiative transfer code.\\
\\In the first part, we consider the contributions of PAHs in the different states (gas-phase and frozen) to the SED, and then, we analyse the effect of disc inclination on the observable PAH feature-to-continuum ratio and the integrated PAH line fluxes.\\
\\Figure \ref{fig:PAH-SED} displays an SED of a protoplanetary disc with PAH emission (\emph{dashed black line}) observed face-on (0$^{\circ}$) that is composed of neutral gas-phase PAH emission (\emph{green}), cationic gas-phase PAH emission (\emph{red}), dust continuum emission (\emph{blue}), and emission from PAHs that are adsorbed on dust grains (\emph{orange}).
Similar to the results of \citet{Maaskant2014}, the PAH emission of our disc model without any dust gaps is dominated by neutral PAHs.
Because gas-phase PAHs undergo the strongest temperature fluctuations, they contribute most to the PAH typical infrared features between $3-13$\,$\mu$m to the spectrum.
In contrast, the adsorbed PAHs are thermally coupled to the dust grains at the radiative equilibrium temperature.
The non-fluctuating thermal emission of the adsorbed PAHs does not significantly contribute to features between $3-9$\,$\mu$m, even though there are 100 times more frozen-out PAHs than gas-phase PAHs in our model and dust grains in the inner disc can reach several hundred Kelvin.
However, there is only contribution to the features at 11\,$\mu$m, 13\,$\mu$m, and 18\,$\mu$m.
For the \emph{smooth-disc} model, the PAH clusters contribute 20\,\% of the emission to the 11\,$\mu$m feature, while the contribution in the \emph{PAH-gap} model is equal between adsorbed clusters and gas-phase PAH.
Consequently, the integrated\footnote{To account for the shift of lines between neutral/cation and broadness of the lines, our integration intervals are 3\,$\mu$m: 3.1-3.4\,$\,\mu$m, 6\,$\mu$m: 6-7\,$\mu$m, and 11\,$\mu$m: 11-12\,$\mu$m.} feature ratios are affected by the disc model.
For the \emph{smooth-disc} model, $I_{3}/I_{11} = 0.3$, $I_{6}/I_{11} = 0.08$, while for the \emph{PAH-gap} model, $I_{3}/I_{11} = 0.1$ and $I_{6}/I_{11} = 0.05$.\\
\\Therefore, the relative intensity between the long-wavelength features and the short-wavelength features does not only resemble the charge state of PAHs \citep[e.g. 6.2\,$\mu$m to 11.3\,$\mu$m ratio][]{Allamandola1999}, but also the relative abundance of PAHs in the disc (hot inner disc versus cold outer disc) and the aggregate form of PAHs (strong temperature fluctuations as monomers in gas-phase versus thermally coupled PAHs and PAH clusters frozen-out on dust grains).
We also find that the total integrated PAH emission over all features in the \emph{gap model} is missing 80\,\% of the \emph{smooth-disc} model PAH luminosity, even though only 15\,\% of the total gas-phase PAHs is additionally adsorbed in the \emph{PAH-gap} model.
Hence, the total PAH emission might not only reflect the total PAH abundance in the disc, but also strongly depends on the spatial distribution of the PAHs within the disc.\\
\\As the next step, we investigated the difference between the \emph{smooth-disc} model and the \emph{PAH-gap} model when the disc is observed under different inclinations.
Figure \ref{fig:PAH_SED_incl} shows the SED of our two disc models for the MIR wavelength range that are accessible to the JWST for different selected disc inclinations. 
For a low and intermediate inclination, the silicate features between 9-12\,$\mu$m contribute significantly to the spectrum, which typically makes the extraction of weak PAH features very difficult because forsterite and enstatite, for example, have peaky features in their opacities \citep{vanBoekel2005}.
Hence, the lowest contamination from the dust grains exists in edge-on discs, which should provide the best PAH spectra that can be obtained.
We wished to analyse only the PAH features, and therefore, we subtracted the continuum from the SED and integrated the residual PAH features in the aforementioned intervals.
We then plot the derived integrated PAH flux ratios for the easiest detectable PAH features at 3\,$\mu$m, 6\,$\mu$m, and 11\,$\mu$m as a function of disc inclination in figure \ref{fig:integrated_lineflux}.
For the models at low to medium inclinations (face-on to 70\,$^\circ$ inclination), the ratio of the integrated line fluxes remains constant, and the inclination does not play a role.
We note again that the difference in $I_{6}/I_{11}$ of the two disc models is related to the different temperature structures of the excited PAHs and not to the level of ionisation because only a very small fraction of the PAH emission originates from ionised molecules.
Only for very high inclinations ($\geq$ 75\,$^\circ$) do all ratios decrease because the short-wavelength PAH emission originating from the inner disc is blocked by the outer disc and can no longer be observed.
As a consequence, the line ratios of the two disc models at high inclinations are barely distinguishable and only provide information on the PAHs of the outer disc.
Even though edge-on discs provide the highest feature-to-continuum ratio and suppress the coinciding silicate feature emission, their SED provides only very limited information about the PAHs and their UV environment in most parts of the disc.

\begin{figure*}
    \centering
    \includegraphics[width=0.99\linewidth]{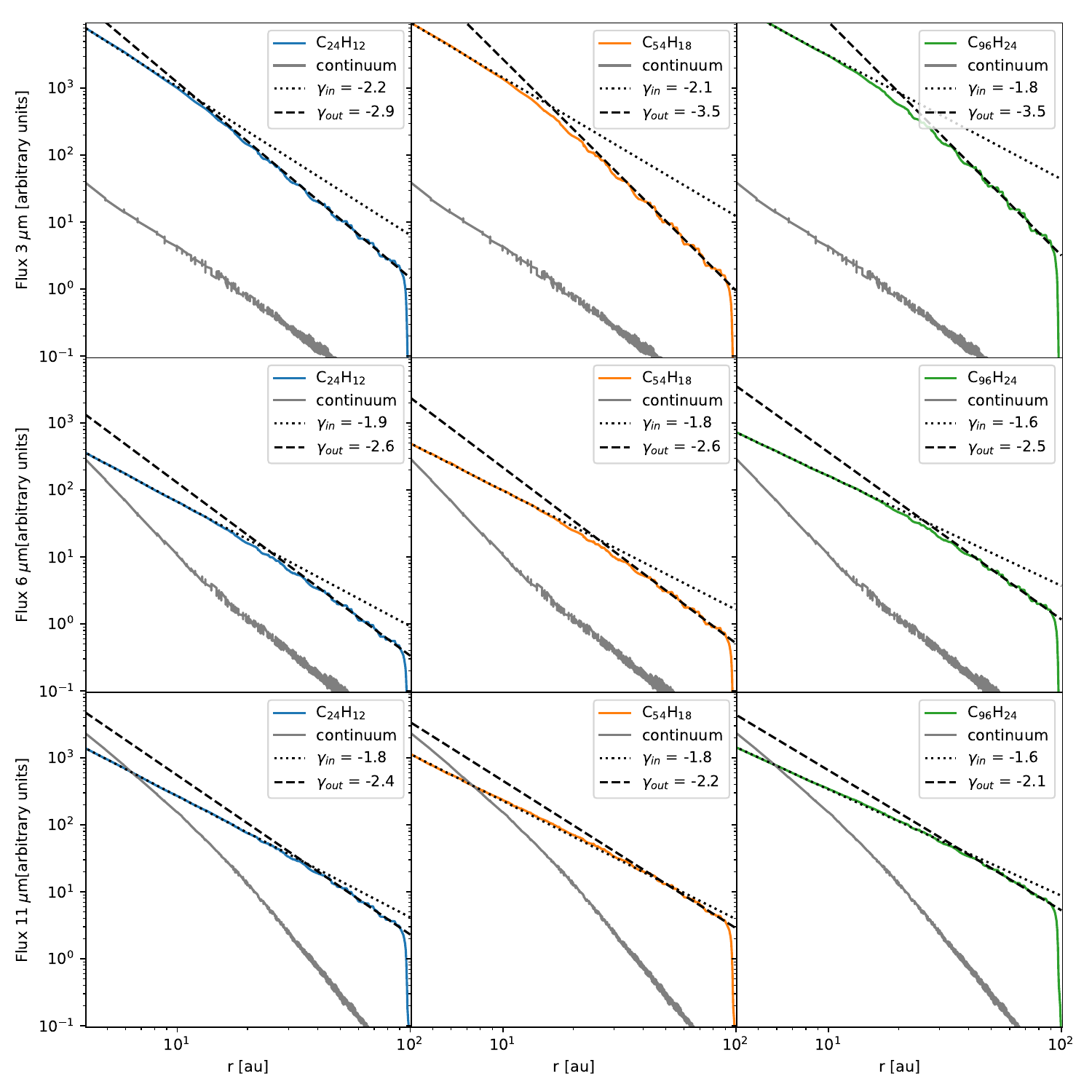}
    \caption{Extracted radial profiles of the most prominent PAH emission features from radiative transfer simulations. The inner and outer disc are fitted with power laws. The inner disc is dominated by multi-photon absorptions with a shallow slope, while the PAHs in the outer disc are heated only through single photons.}
    \label{fig:radial_powerlaws_all}
\end{figure*}

\begin{figure*}
    \centering
    \includegraphics[width=0.99\linewidth]{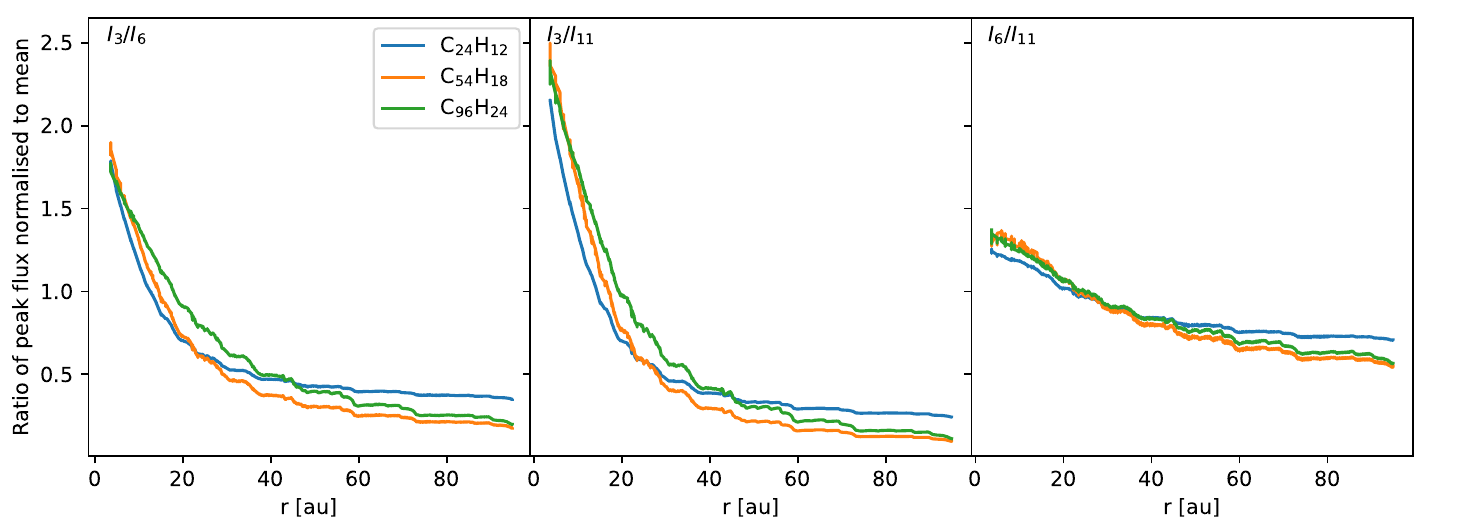}
    \caption{Ratio of the most prominent PAH line ratios normalised to the mean value obtained from the whole disc. Because the 3\,$\mu$m line is very sensitive to the UV field of the disc, the line ratio changes very strongly throughout the disc. However, the size of the PAH molecule plays only a minor role.}
    \label{fig:ratio_variation}
\end{figure*}

\subsection{Radial profiles of the PAH emission}
We furthermore used RADMC-3D to produce artifical images to determine the radial profiles of the PAH emission at the most prominent emission features that will become accessible through the JWST.
We always produced two artificial images: one including PAH emission, and the other containing only continuum emission produced by the dust grains so that we were able t calculated the contribution from the PAHs by subtracting the continuum for every pixel.
We produced our images with a spatial resolution of 2.3\,au/px at the feature wavelengths for coronene C$_{24}$H$_{12}$ at 3.259\,$\mu$m, 6.277\,$\mu$m, and 11.59\,$\mu$m, for circumcoronene C$_{54}$H$_{18}$ at 3.26\,$\mu$m, 6.26\,$\mu$m, and 11.1\,$\mu$m, and for circumcircumcoronene C$_{96}$H$_{24}$ at 3.26\,$\mu$m, 6.32\,$\mu$m and 11.0\,$\mu$m. 
As the fraction of ionised emission is minor, we did not need to integrate over a wavelength interval around the feature to take the shift in the line wavelength into account. 
We then determined the distance of each pixel to the centre of the image and plot the flux in each pixel at the corresponding distance.\\
\\Figure \ref{fig:radial_powerlaws_all} shows the radial profiles for all three PAH molecules (\emph{coloured lines}) and continuum emission (\emph{grey}).
The radial profiles of the PAH emission between the inner and outer disc differ for all molecules, and we find that each region can be described by a simple power law with a shape of $f(x)=Ax^\gamma$.
We attribute the difference between inner and outer disc to the temperature fluctuations of the corresponding PAHs.
In the inner disc, the temperature of the PAHs fluctuates in a narrow distribution around the radiative equilibrium temperature in a multi-photon regime, and all PAH features between $3-11$\,$\mu$m are excited because the temperature is high enough to allow emission the entire time.
The shallow power-law slope therefore reflects multiple properties.
On the one hand, the radial density gradient in the number of emitting PAHs, as well as the flaring angle of the disc, because more flaring allows more absorption of UV photons at a given distance.
On the other hand, the width of the temperature probability distribution, where a broader distribution leads to a slightly steeper radial gradient (the slope is more shallow for larger PAHs).
In the outer disc, PAHs are mostly flash-heated by single photons and then cool down to the background temperature.
During cooling, the excitation temperature is high enough most of the time to lead to emission of PAH features at 6\,$\mu$m and 11\,$\mu$m.
However, the 3\,$\mu$m feature requires at least two photons to be excited for an extended period of time, and the emission of the line depends on the high-energy tail of the temperature fluctuations. 
Therefore, the larger the PAH, the greater the heat capacity and the shorter the period of time for which the 3\,$\mu$m feature is excited.
As a consequence, the power law becomes steeper for larger PAHs.
Additionally, we note that the transition between the two power laws increases with wavelength of the emission feature and size of the PAH, even though the transition at 6\,$\mu$m is very similar for all three PAHs.\\
\\Finally, figure \ref{fig:ratio_variation} shows the relative peak intensities $I_\text{peak}$ of the three features normalised to the mean value that is obtained from the whole disc. 
As expected, the $I_3$/$I_{11}$ has the strongest relative variation through the disc, showing a variation of a factor 25 between the inner and outer disc.
The $I_3$/$I_{6}$ intensities behave very similarly, but show a weaker variation of only a factor 10.
Because both the $I_6$ and $I_{11}$ intensities are not as sensitive to the temperature fluctuations, the variation throughout the disc is only minor, of a factor of 2.
We do not find a relevant change in the variation of intensities for different PAH species and sizes.


\section{Discussion}
Our results show that even for our two very simple radiative transfer models, the information that can be obtained through investigation of the SED is very limited.
In the literature, the observed spread in PAH luminosities in Herbig stars ranges from $L_\text{PAH}/L_* = 8.3\cdot 10^{-5}$ (HD\,58647) to $L_\text{PAH}/L_* = 2.6 \cdot 10^{-2}$ (Oph\,IRS\,48) \citep{Acke2010}. This difference has been analysed and interpreted as a difference in PAH mass between $0.052 \cdot 10^{-6} M_\oplus$ and $24.3 \cdot 10^{-6} M_\oplus$ \citep{Seok2017}.
This correspsonds to an almost linear relation between PAH luminosity and abundance.
Because our simple \emph{PAH-gap} model lacks only 15\% of the PAHs compared to the \emph{smooth-disc} model, but has 80\% less PAH emission, it shows that the correlation between PAH abundance and PAH luminosity is not necessarily linear.
This might imply that the observed spread in PAH abundance might be smaller than previously calculated and might instead indicate a different spatial distribution of PAHs in the various discs.
Especially the various different substructures of simple carbon species seen with ALMA in the MAPS survey \citep[e.g. see][]{Oberg2021, Law2021} links the majority, but not all, substructures to the dust continuum and the chemical interaction and evolution between the disc midplane and surface.
Hence, given the complexity and amount of substructures seen with ALMA in the dust continuum \citep[e.g. with DSHARP,][]{Andrews2018}, at least some information about the distribution of PAHs and substructures in the disc must be known or assumed in order to derive meaningful PAH abundances because their chemical evolution is likely also linked to dust grains \citep{Lange2023}.
Additionally, information about the approximate inclination ($i\geq 70^\circ$ or not) of the disc is useful to interpret the extracted information from an SED correctly.
The JWST facilities and IFU of NIRspec and MIRI will provide this information in the near future.\\
\\We furthermore wished to evaluate the impact of the temperature structure on the measured relative line fluxes between 3\,$\mu$m, 6\,$\mu$m and 11\,$\mu$m. 
Focussing on the example of coronene C$_{24}$H$_{12}$, the intrinsic integrated cross section of the 11\,$\mu$m feature is $\sigma_{11}=180$\,km/mol (neutral) and $\sigma_{11}=190$\,km/mol (cation), while at 6.2\,$\mu$m, the integrated cross section is $\sigma_{6}=13$\,km/mol (neutral) and $\sigma_{6}=455$\,km/mol (cation).
Hence, the change in intrinsic cross section through charging of the PAH is a factor of 35 for the $\sigma_{6/11}$ ratio.
In our radial profiles ,however, we only see a variation of $I_{6/11}$ throughout the disc of a factor of 2, which arises from the different temperature fluctuations.
The exact variation of the intrinsic cross section also depends on the considered molecule and can be much larger when circumcircumcoronene C$_{96}$H$_{24}$ is considered, for instance.
Nevertheless, due to the insensitivity to the temperature structure of the disc, the $I_6/I_{11}$ line ratio is a very good tracer of the ionisation state at all distances in the disc, allowing for more detailed electron chemistry models.\\
\\Based on the intrinsic cross sections for coronene, we expect $ 0.07 \geq I_6/I_{11} \geq 2.5$.
However, when we compare the observed ratios for $I_6/I_{11}$ with Spitzer \citep{Acke2010}, which typicall lies at $ 1 \geq I_6/I_{11} \geq 3$, our simple disc models are not able to reproduce these values because they heavily underestimate the fraction of ionised emission through usage of a smooth disc without dust gaps.
ALMA observations show that most of the discs have prominent substructures. This emphasises the need of a detailed model for each individual disc in order to explain the observed feature ratios  \citep[see also][]{Maaskant2014}.\\
\\When we compare for coronene C$_{24}$H$_{12}$ the 3\,$\mu$m and 11\,$\mu$m cross sections, the intrinsic integrated cross-section ratio is $\sigma_{3/11} = 0.8$ (neutral) and $\sigma_{3/11} = 0.3$ (cation).
Even though the 3\,$\mu$m feature is suppressed for cationic emission,
the observed change in the intensity ratio in our models is a factor of 35.
Hence, the $I_3/I_{11}$ is clearly a strong indicator for the UV field rather than the ionisation state because the 3\,$\mu$m is very sensitive to the temperature fluctuations.
Therefore, we propose to use the 3\,$\mu$m feature to trace back the UV field.
The radial profiles of all three PAH features might be used to characterise the UV field by determining the position at which the transition from the multi-photon to single-photon regime occurs.
However, detailed modelling of the disc with its structure is likely necessary to obtain reliable results, and further studies of the sensitivity of the transition on the disc and stellar spectrum are required.


\section{Conclusions}
We have simulated the SED of two protoplanetary discs: one disc with a constant abundance of PAHs, and the other disc where the inner disc has 10\% fewer PAHs than the outer disc. We furthermore provided radial profiles for the 3\,$\mu$m, 6\,$\mu$m, and 11\,$\mu$m emission.
The strength of the integrated PAH emission barely reflects the PAH abundance, but is rather sensitive to the spatial distribution of the PAHs and their absorbed UV radiation. Considering the complex substructures of protoplanetary discs, it is questionable how conclusive derived PAH abundances from SEDs are without a proper consideration of the gaps, rings, and flaring angle of the disc.
The radial power-law slope of the PAH emission provides information about the size of the PAHs in a disc because especially the 3\,$\mu$m feature is very sensitive to temperature fluctuations. The transition from a multi-photon regime to a single-photon absorption regime provides information about the underlying UV field at the transition location of the radial profiles.
However, for quantitative statements about the PAH abundance, the PAH size, and the underlying UV field, a detailed model of the corresponding disc is required, which needs to be supplemented by additional observations such as observations with ALMA and SPHERE.

\begin{acknowledgements}
 The authors thank the anonymous referee. The authors further thank Alessandra Candian for very fruitful and useful discussions. K.L. acknowledges funding from the Nederlandse Onderzoekschool Voor Astronomie (NOVA) project number R.2320.0130. C.D. acknowledges funding from the Netherlands Organisation for Scientific Research (NWO) TOP-1 grant as part of the research program “Herbig Ae/Be stars, Rosetta stones for understanding the formation of planetary systems”, project number 614.001.552. 
Studies of interstellar PAHs at Leiden Observatory are supported through a Spinoza award from the Dutch research council, NWO.
\end{acknowledgements}

\bibliographystyle{aa} 
\bibliography{library.bib} 

\appendix

\end{document}